\documentclass[aps,prl,twocolumn,superscriptaddress]{revtex4}
\usepackage{graphicx}
\begin{document}


\title{The anomalous behavior of coefficient of normal restitution in 
the oblique impact}


\author{Hiroto Kuninaka}
\email[E-mail: ]{kuninaka@kuchem.kyoto-u.ac.jp}
\thanks{Present address: Department of Chemistry, Kyoto University, 
Sakyo-ku, Kyoto, Japan, 606-8502}
\affiliation{Graduate School of Human and Environmental Studies, 
Kyoto University, Sakyo-ku, Kyoto, Japan, 606-8501}
\author{Hisao Hayakawa}
\affiliation{Department of Physics, Yoshida-south campus, 
Kyoto University, Sakyo-ku, Kyoto, Japan, 606-8501
 }

\date{\today}

\begin{abstract}
The coefficient of normal restitution in an oblique impact is
theoretically studied. 
Using a two-dimensional lattice model for an elastic disk 
and an elastic wall, 
we demonstrate
that the coefficient of normal restitution can exceed unity 
and has a peak against the incident angle 
in our simulation. 
We also explain this behavior based upon 
a phenomenological theory.
\end{abstract}


\maketitle

The coefficient of normal restitution $e$ 
is introduced to determine the normal component 
of the post-collisional velocity in the collision 
of two materials. 
The coefficient $e$ is defined by 
\begin{equation}\label{COR}
{\bf v}(\tau_{c}) \cdot {\bf n}=-e {\bf v}(0) \cdot {\bf n},
\end{equation} 
where ${\bf v}(\tau)$ is the relative velocity of 
the centers of mass of two colliding materials 
at time $\tau$ measured from the initial contact, 
$\tau_{c}$ is the duration of a collision, 
and ${\bf n}$ is the unit vector normal to the contact plane. 
Though some text books of elementary physics state 
that $e$ is a material constant, 
many experiments and simulations show that $e$ decreases 
with increasing impact velocity\cite{stronge}. 
The dependence of $e$ on the low impact velocity is theoretically 
treated by the quasi-static theory
\cite{kuwabara,morgado,brilliantov96}. 
We also recognize that $e$ can be less than unity 
for the normal impact without the introduction of any explicit 
dissipation, because the macroscopic inelasticities 
originate in the transfer of the energy 
from the translational mode to the internal modes 
such as the vibrations\cite{morgado,gerl,ces}. 
 
While $e$ has been believed to be less than unity 
in most situations, 
it is recently reported that $e$ can exceed unity 
in oblique impacts\cite{smith,calsamiglia,louge}. 
In particular, Louge and Adams\cite{louge} observed 
oblique impacts of a hard aluminum oxide sphere 
on a thick elastoplastic polycarbonate plate 
and found that $e$ grows monotonically with the magnitude of 
the tangent of the incident angle $\gamma$. 
In their experiment, 
Young's modulus of the plate is $100$ times smaller 
than that of the aluminum oxide sphere. 
They also suggested that $e$ can exceed unity 
for the most oblique impacts. 
 
In this letter, 
we demonstrate 
that our two-dimensional simulation of the oblique impact 
based on Hamilton's equation 
has yielded an increasing $e$ with $\tan\gamma$ and 
$e$ exceeds unity at the critical incident angle. 
Finally, we explain our results 
by our phenomenological theory. 

\begin{figure}[htbp]
\begin{center}
\includegraphics[width=0.5\textwidth]{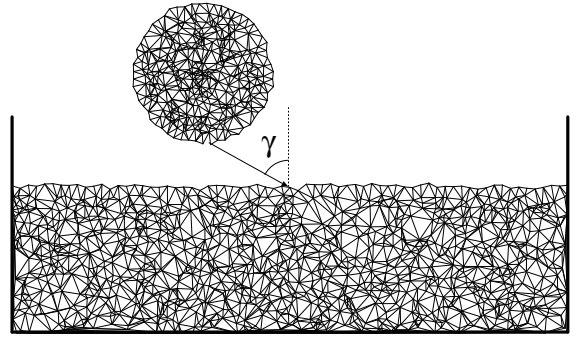}
\end{center}
\caption{The elastic disk and wall consisted of random lattice.}
\label{fig1}
\end{figure}

Let us introduce our numerical model\cite{kuninaka_jpsj2003}. 
Our model consists of an elastic disk and an elastic wall 
(Fig. \ref{fig1}). 
The width and the height of the wall are $8 R$ and $2 R$, 
respectively, where $R$ is the mean radius of the undeformed disk. 
The both side ends and the bottom of the wall are fixed. 
We place $800$ mass points at random in a disk with the radius $R$ 
and $4000$ mass points at random in a wall 
for the disk and the wall, respectively.  
We connect each mass point with its neighbor mass points 
by the Delaunay triangulation algorithm\cite{delaunay}, 
and undeformed nonlinear springs are placed on all the connections.


Each mass point $i$ on the lower half boundary of the disk 
feels the force, 
${\bf F}(l_{s}^{(i)})=aV_0\exp(-a l_{s}^{(i)}){\bf n}_{s}^{(i)}$, 
where $l_{s}^{(i)}$ is the distance 
between $i$-th surface mass point of the disk and 
the nearest surface spring of the wall, $a=300/R$, $V_0=amc^2R/2$, 
$m$ is the mass of each mass point, $c$ is the one-dimensional 
velocity of sound, 
and ${\bf n}_{s}^{(i)}$ is the unit vector normal to the connection 
between two surface mass points of the wall
\cite{kuninaka_jpsj2003}. 
We should note that the strong repulsion ${\bf F}(l_{s}^{(i)})$ 
is introduced to inhibit the penetration of the disk 
to the surface of the wall\cite{gerl}. 
Thus, the dynamical equation of motion for each mass point $i$ 
of the lower half boundary of the disk is described by
\begin{equation}\label{keq}
m \frac{d^{2}{\bf r}_{i}}{d \tau^{2}}=
\sum^{N_{i}}_{j=1} \left\{-k_{a} {\bf x}_{ij} - k_{b} {\bf x}_{ij}^{3}
\right\}+\Theta(l_{th}-l_{s}^{(i)}) a V_0 \exp(-a l_{s}^{(i)})
{\bf n}_{s}^{(i)},
\end{equation}
where ${\bf r}_{i}$ is the position of $i$-th mass point, 
$N_{i}$ is the number of mass points connected to $i$-th mass point, 
${\bf x}_{ij}$ is the relative deformation of the spring 
between $i$-th and $j$-th connected mass points, 
$k_{a}$ and $k_{b}=k_{a}\times10^{-3}/R^{2}$ are the spring constants. 
Here we introduce the step function $\Theta(x)$, 
i.e. $\Theta(x)=1$ for $x \ge 0$ and $\Theta(x)=0$ for $x < 0$, 
and the threshold length $l_{th}$ which is the average of 
the natural lengths of the springs of the disk. 
For internal mass points, the last term of 
the right hand side of eq.(\ref{keq}) is omitted. 
In most of our simulations, 
we adopt $k_{a} = k^{(d)}_{a} =1.0 \times m c^2/R^2$ 
for the disk and 
$k_{a} = k^{(w)}_{a} =1.0 \times 10^{-2} m c^2/R^2$ for the wall. 
We do not introduce any dissipative mechanism in this model. 
Thus, during a collision, 
a part of initial translational energy of the disk is 
distributed into the vibrational energy of the disk and the wall. 
It should be noted that the macroscopic dissipation 
can be interpreted as the irreversible transfer of the energy 
from the translational motion to the internal vibration. 
When we introduce explicit dissipations in the model 
and add the gravity to the disk, we have confirmed 
that the compression of the disk can be described 
by two-dimensional Hertzian contact theory
\cite{kuninaka_doctor}.  

In this model, the roughness of the surfaces is important 
to make the disk rotate after a collision\cite{kuninaka_jpsj2003}. 
We modify the positions of the surface mass points of the flat 
wall and the smooth disk by using normal random numbers 
whose average and standard deviation are 
$0$ and $3 \times 10^{-2}R$, respectively. 

Poisson's ratio $\nu$ and Young's modulus $E$ of this model 
can be evaluated by adding the external force to 
stretch the rectangle of random lattice numerically. 
We obtain Poisson's ratio $\nu=(7.50 \pm 0.11) \times 10^{-2}$ 
and Young's modulus $E=(9.54 \pm 0.231)\times 10^{3}mc^{2}/R^{2}$, 
respectively\cite{kuninaka_jpsj2003}. 

We solve the dynamical equation of motion (\ref{keq}) 
for each mass point with the initial speed $|{\bf v}(0)|=0.1 c$ 
and the incident angle $\gamma$,  
and determine $e$ for each $\gamma$ according to eq.(\ref{COR}). 
All the results in this letter are obtained by averaging 
the results of 100 disks with different configurations of 
mass points. 
We use the fourth order symplectic integrator 
with the time step $\Delta \tau = 10^{-3}R/c$. 

\begin{figure}[htbp]
\begin{center}
\includegraphics[width=0.4\textwidth]{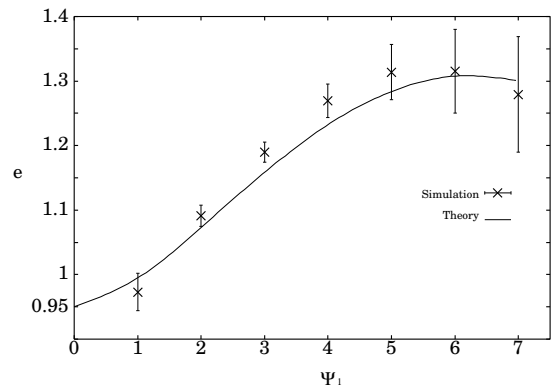}
\end{center}
\caption{Numerical and theoretical results of the relation between
 $\Psi_1$ and $e$.}
\label{fig2}
\end{figure}

Figure \ref{fig2} is the relation between $e$ and 
$\Psi_1 \equiv -({\bf v}(0)\cdot{\bf t})/({\bf v}(0)\cdot{\bf n})
=\tan\gamma$, 
where ${\bf t}$ is the unit vector vertical to ${\bf n}$. 
The cross point is the mean value and the error bar is  
the standard deviation of 100 samples for each $\gamma$. 
This result shows that $e$ increases with increasing $\Psi_1$ 
to exceed unity, and has a peak around $\Psi_1 = 6.0$. 
The behavior of $e$ having the peak is contrast to 
that in the experiment by Louge and Adams\cite{louge}. 

Here, let us explain 
our results. 
Louge and Adams\cite{louge} suggest 
that their results can be explained by the rotation of 
the normal unit vector ${\bf n}$ arising from 
the local deformation of the wall's surface. 
Thus, we aim to determine the angle of rotation of the unit vector 
$\alpha$ at each $\gamma$ from the theory of elasticity. 

\begin{figure}[htbp]
\begin{center}
\includegraphics[width=0.4\textwidth]{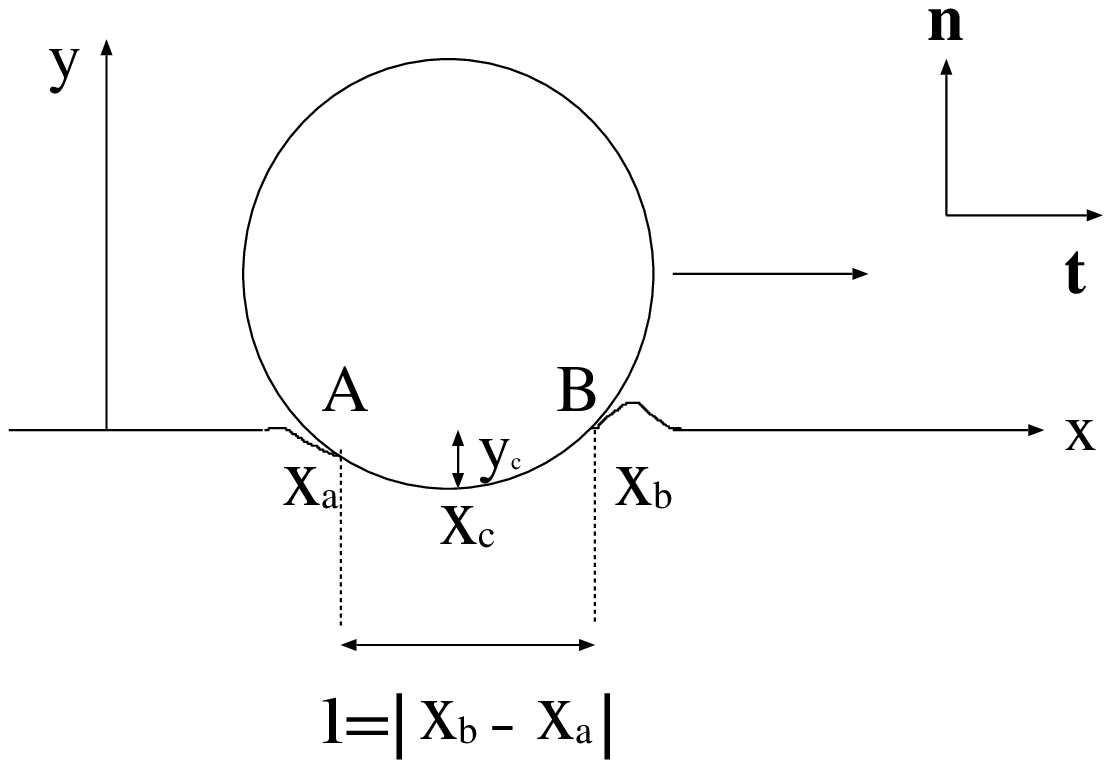}
\end{center}
\caption{The schematic figure of a hard disk sliding on a soft wall. 
$x$ coordinates of both ends of the contact area AB are $x=x_a$ and 
$x=x_b$.}
\label{fig3}
\end{figure}

\begin{figure}[htbp]
\begin{center}
\includegraphics[width=0.4\textwidth]{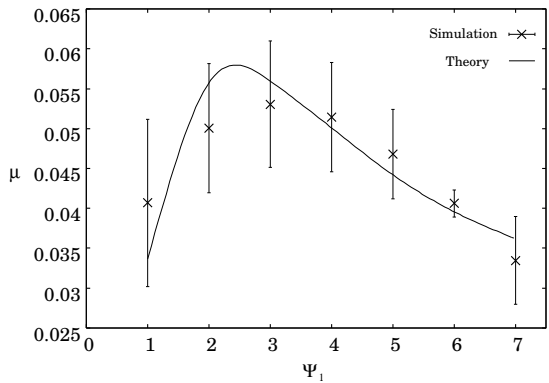}
\end{center}
\caption{Numerical and theoretical results of the relation between
 $\Psi_1$ and $\mu$.}
\label{fig4}
\end{figure}

Figure \ref{fig3} is the schematic figure of a hard disk moving 
from the left to the right on a wall,  
where $l\equiv|x_b-x_a|$. 
Assuming that $l$ is small compared to $R$, 
the perimeter of the contact is approximated by 
\begin{equation}\label{contact}
 f(x)=(x-x_c)^2/2R-y_c,
\end{equation}
where $(x_c,y_c)$ is the lowest position of the disk. 
To calculate $\tan\alpha \equiv (f(x_b)-f(x_a))/l$, 
we need to know the ratio of $|x_c-x_a|$ to $l$.  
From the theory of elasticity\cite{galin}, 
this ratio can be estimated as 
\begin{equation}\label{ratio1}
 \frac{x_c-x_a}{l} = 1-\theta \hspace{5mm} \text{with} \hspace{5mm}
 \theta = \frac{1}{\pi}\arctan \frac{1-2\nu}{\mu (2-2\nu)},
\end{equation}
where $\mu$ is the coefficient of the friction. 
We evaluate $\mu$ 
by $\mu \equiv |{\bf J} \cdot {\bf t}|/|{\bf J} \cdot {\bf n}|$ 
with ${\bf J}=M({\bf v}(\tau)-{\bf v}(0))$, where $M$ is the mass 
of the disk. 
The cross points in Fig.\ref{fig4} represent $\mu$ calculated 
from our simulation, where $\mu$ has a peak 
around $\Psi_1=3.0$. 
From eqs.(\ref{contact}) and (\ref{ratio1}),  
$\tan\alpha$ is given by 
\begin{equation}\label{tanalpha2}
\tan \alpha = \frac{2\theta-1}{2-2\theta} \frac{|x_c-x_a|}{R},
\end{equation}
where we evaluate $|x_c-x_a|=0.55R$ 
from the maximum value of the compression of the disk. 
From eqs.(\ref{ratio1}) and (\ref{tanalpha2}), 
we obtain the relation between $\Psi_1$ and $\tan\alpha$.

Next, let us calculate $e$ 
from the relation between $\Psi_1$ and $\tan\alpha$. 
We introduce the rotated unit vectors, 
${\bf n}_{\alpha}$ and ${\bf t}_{\alpha}$, as 
${\bf n}_{\alpha} = \cos\alpha {\bf n}-\sin\alpha {\bf t}$ and 
${\bf t}_{\alpha} = \sin\alpha {\bf n}+\cos\alpha {\bf t}$, 
respectively. 
By introducing 
$e_{\alpha} \equiv -({\bf v}(\tau_{c}) \cdot {\bf n}_{\alpha})/({\bf v}(0) \cdot
{\bf n}_{\alpha})$,  
we can express $e$ in terms of $e_{\alpha}$ as 
\begin{equation}\label{CORmod2}
e=\frac{e_{\alpha}+\Psi_2^{\alpha} \tan\alpha}{1-\Psi_1^{\alpha} \tan\alpha},
\end{equation}
where 
$\Psi_1^{\alpha}=-({\bf v}(0) \cdot {\bf t}_{\alpha})/({\bf v}(0) \cdot 
{\bf n}_{\alpha})$ 
and $\Psi_2^{\alpha}=-({\bf v}(\tau_{c}) \cdot {\bf t}_{\alpha})/({\bf v}(0) \cdot 
{\bf n}_{\alpha})$. 
$\Psi_1^{\alpha}$ also can be rewritten as 
\begin{equation}\label{psi1}
\Psi_{1}^{\alpha}=(\Psi_1-\tan\alpha)/(1+\Psi_1 \tan\alpha).
\end{equation}
On the other hand, in the oblique impact of slipping disks, 
$\Psi_2^{\alpha}$ is given by  
\begin{equation}\label{from_wal}
\Psi_2^{\alpha}=\Psi_1^{\alpha}-3(1+e_{\alpha})\mu_{\alpha}
\end{equation}
in the two-dimensional situation\cite{walton_rheol}.  
In eq.(\ref{from_wal}), $\mu_{\alpha}$, defined by 
$\mu_{\alpha}=|{\bf J} \cdot {\bf t}_{\alpha}|/|{\bf J} \cdot {\bf
n}_{\alpha}|$, is given by 
\begin{equation}\label{mualpha}
\mu_{\alpha}=\frac{\mu+\tan\alpha}{1-\mu\tan\alpha}.
\end{equation}
To draw the solid line in Fig.\ref{fig2}, at first, 
we calculate $\mu$ and $\tan\alpha$ for each $\Psi_1$. 
By choosing a fitting parameter $e_{\alpha}=0.95$ and 
substituting eqs.(\ref{psi1})-(\ref{mualpha}) into eq.(\ref{CORmod2}) 
we obtain $e$ as a function of $\alpha$, $\mu$, and $\Psi_{1}$. 
All points are interpolated with the cubic spline interpolation method 
to draw the solid curve. 
Such the theoretical description of $e$ 
is qualitatively consistent with our numerical result, 
though the theoretical value is a little smaller 
than the observed value. 

Now, let us consider how $\mu$ depends on $\Psi_1$ 
based on a phenomenological argument. 
For simplicity, we replace the roughness on the surface 
of the wall by a periodic array of asperities.
When the disk hits one asperity,  
 a fraction of the energy is absorbed in the wall. We assume that 
the tangential velocity $v_t^{(i-1)}$ changes to
$v_t^{(i)}=(1-\eta)v_t^{(i-1)}$ when the disk hits the $i$-th asperity.
Thus, if the disk contacts $N$ asperities during the impact, the
tangential speed at the release point 
becomes $v_t(\tau_{c})=v_{t}^{(N)}=(1-\eta)^Nv_t(0)$.
Here we estimate the number of contacted asperities during the
collision as $N=\rho l_{sl}$, where $\rho$ is the number of 
the asperities 
in a unit length on the surface and $l_{sl}$
is the length of sliding  which can be evaluated as 
$l_{sl}=v_t(0)\tau_{c}$ with $\tau_{c}=\pi (R/c)
\sqrt{\ln\left(4 c/v_n(0)\right)}$\cite{ces}.
Thus, the tangential impulse $J_t\equiv {\bf J}\cdot {\bf t}$ 
is approximated by
\begin{equation}
J_t=M\{(1-\eta)^{\rho l_{sl}}-1\}v_t(0)\simeq -\eta M \rho l_{sl} 
v_t(0),
\end{equation}
where we assume small $\eta$. 
Now, we should answer the question how to determine $\eta$.
When the impact velocity is large enough, $\eta$ becomes smaller 
because the asperities are broken down when the disk hits them.
Therefore we may assume the form 
$\eta=\eta_0/(1+\beta v^{2}_{t}(0)/v^{2}_{n}(0))
=\eta_0/(1+\beta \Psi^{2}_{1})$ with 
the introduction of the dimensionless parameters $\eta_0$ and $\beta$.
Taking into account 
$J_n \equiv {\bf J}\cdot{\bf n}=-M(1+e)v_n(0)$ and the
definition of $\mu=J_t/J_n$ we obtain
\begin{equation}
\mu=\frac{\pi\eta_{0} \rho R}{1+e}
\frac{\Psi_{1}}{1+\beta \Psi^{2}_{1}}
\sqrt{\frac{\Psi^{2}_{1}}{1+\Psi^{2}_{1}}
\ln\left(40\sqrt{1+\Psi^{2}_{1}}\right)} \frac{|{\bf v}(0)|}{c}.
\label{toy}
\end{equation}
Here we use $\cos\gamma=\sqrt{1/(1+\Psi^{2}_{1})}$,
$\sin\gamma=\sqrt{\Psi^{2}_{1}/(1+\Psi^{2}_{1})}$. 
We adopt the numerical results for the value of $e$ 
at each $\Psi_1$ (Fig.\ref{fig2}) and $|{\bf v}(0)|/c = 0.1$. 
The solid curve in Fig.\ref{fig4} is eq.(\ref{toy}) 
with fitting parameters $\beta=0.21$ and $\eta_0 \rho R=0.18$, 
which reproduces our numerical result. 
We do not claim that our simple argument explains 
the experimental result because of two fitting parameters. 
However, we would like to emphasize that our picture 
captures the essence of the behavior of $\mu$ in the oblique impact.

Let us discuss our result. 
First, we emphasize that the novel phenomena of $e$ exceeding unity 
are obtained from the local deformation of the soft wall 
for the oblique impacts of a hard disk. 
When we simulate the impact between a disk and a hard wall,  
for $k^{(w)}_a = 10 \times k^{(d)}_a$, 
$e$ fluctuates around a constant to exceed unity abruptly 
around $\Psi_1=4.5$\cite{kuninaka_doctor}. 
This tendency resembles the experimental results 
by Calsamiglia {\it et al.}\cite{calsamiglia}. 
Thus, for smooth increase of $e$ to exceed unity, 
the wall should be softer than the disk. 
In addition, 
it is important to fix the initial kinetic energy of the disk. 
So far, we have confirmed that $e$ does not exceed unity when $\Psi_1$ is 
controlled by changing $v_t$ with fixed
$v_n$\cite{kuninaka_jpsj2003}. 

The second, the initial velocity of the disk and the local deformation 
of the wall are much larger than those in the experimental ones 
in ref.\cite{louge}. 
They cause the most significant difference 
between our result and their result\cite{louge}.  
Because of the high speed impact in our case, 
there is a peak of $e$ for small $\gamma$. 
In fact, our simulation with 
$|{\bf v}(0)|=0.01c$ shows the shift of the peak for larger $\gamma$. 
Therefore, we expect that our model reproduces the result of 
ref.\cite{louge} for the low impact speed. 
In addition,  we have carried out simulations 
with a disk of $400$ mass points and a wall of $2000$ mass points 
to check the effect of the system size. 
Although there is a slight difference between the results, 
the data are also well reproducible by our phenomenological theory.  

The third, the local deformation of the wall also affects 
the relation between $\mu$ and $\Psi_1$. 
In early studies, 
it has been shown that $\mu$ depends on the impact velocity
\cite{gorham,louge}. 
In our simulation, $\mu$ has a peak around $\Psi_1=3.0$. 
This behavior is interpreted as 
that the asperities are flattened for large $v_t$. 
Equation (\ref{toy}) indicates that $\mu$ can increase 
with increasing $\gamma$ as in ref.\cite{louge} 
if we choose a suitable set of $\beta$ and $\eta_{0} \rho$. 
The difference between the results of their experiment and our
simulation may be explained by the choice of these parameters.

The fourth, we adopt the static theory of elasticity to explain our
numerical results in this letter. 
However, it is important to solve the time-dependent equation 
of the deformation of the wall's surface 
to analyze the dynamics of impact phenomena. 
The dynamical analysis will be our future task.

In final, we indicate that the friction coefficient is derived from
our Hamiltonian model. This friction comes from the irreversible 
energy transfer of the macroscopic translational motion 
to the internal vibration. The irreversibility is indeed related 
to the second law of thermodynamics 
if the number of the internal degrees of freedom is infinite. However, 
the irreversibility in our system which includes only $10^3$ mass points
in a disk is incomplete as we can see that
the Hertzian contact theory cannot be recovered 
without introduction of the explicit dissipation. 
We believe that the complete treatment of inelastic collision
of macroscopic materials without introduction of dissipations will
be a fundamental subject of nonequilibrium statistical mechanics.

In summary, we have carried out the two-dimensional simulation 
of the oblique impact of an elastic disk on an elastic wall. 
We have found that  $e$ can exceed unity in the oblique impact, 
which is attributed to the local deformation of the wall. 
We have estimated the magnitude of the local defomation $\alpha$ 
based on the static theory of elasticity and derived the relation 
between $e$ and $\Psi_1$ 
by taking into account the rotation of the normal unit vector of 
the wall's surface. The relation between $\mu$ and $\Psi_1$ is also 
related to the local deformation and is explained 
by the simple argument. 

\begin{acknowledgments}
We would like to thank Y. Tanaka and S. Nagahiro for their 
valuable comments. 
We would also like to thank M. Y. Louge for his detailed 
introduction of their experiment to us and some valuable comments. 
Parts of numerical computation in this work were carried out 
at Yukawa Institute Computer Facility. 
This study is partially supported by the Grant-in-Aid of 
Ministry of Education, Culture, Sports, Science and Technology (MEXT), 
Japan (Grant No. 15540393) and the Grant-in-Aid for the 21st century COE 
"Center for Diversity and Universality in Physics" from MEXT, Japan. 
\end{acknowledgments}

\bibliography{impact}

\end{document}